\documentclass[aps,twocolumn]{revtex4}
\usepackage{dcolumn}
\usepackage{graphicx}
\usepackage{amsmath}
\usepackage{amsfonts}
\usepackage{amssymb}
\usepackage{psfrag}
\usepackage{wrapfig}
\usepackage{subfigure}
\usepackage{makeidx}
\usepackage{bm}
\usepackage{epsf}
\usepackage{multirow}
\usepackage[colorlinks,urlcolor=blue,citecolor=blue]{hyperref}
\usepackage{amsthm,amsmath,amssymb}
\usepackage{mathrsfs}

\begin{document}
\title{Dirac  Monopoles in Rogue Waves}
\author{L.-C. Zhao$^{1,2,3}$}
\author{L.-Z. Meng$^{1}$}
\author{Y.-H. Qin$^{1}$}
\author{Z.-Y. Yang$^{1,2,3}$}
\author{J. Liu$^{4}$ }
\email{jliu@gscaep.ac.cn}
\address{$^{1}$School of Physics, Northwest University, Xi'an 710127, China}
\address{$^{2}$Peng Huanwu Center for Fundamental Theory, Xi'an 710127, China}
\address{$^{3}$Shaanxi Key Laboratory for Theoretical Physics Frontiers, Xi'an 710127, China}
\address{$^{4}$Graduate School, China Academy of Engineering Physics, Beijing 100193, China}
\begin{abstract}
We for the first time demonstrate that the widely existed nonlinear waves such as rogue waves,  contain Dirac  monopoles. We find that the density zeros of these nonlinear waves on an extended complex plane can constitute the Dirac virtual magnetic monopole fields with a quantized flux of elementary $\pi$. We then can explain the exotic property of ``appearing from nowhere and disappearing without a trace" of rogue waves by means of a Dirac  monopole collision mechanism. The maximum amplification ratio and multiple phase steps of high-order rogue waves are found to be closely related to the number of their contained  monopoles. Important implications of the intrinsic  virtual magnetic monopole fields are discussed.
\end{abstract}
\date{\today}

\maketitle

\emph{Introduction}---Magnetic monopole is a hypothetical elementary particle representing an isolated source of magnetic field with only one magnetic pole (N without S, or vice versa) and a quantized magnetic charge \cite{MMr,MMr2}. Dirac first  demonstrated that a virtual monopole is associated with nodal singularity of wave function and can be described well by topological vector potential with a line singularity \cite{Dirac}.
 The concept of Dirac monopoles provide a fertile playground for theoretical ideas \cite{Rebbi} and are also relevant to physical considerations as they are necessarily present in all grand unified models \cite{Hooft,sff}.  Although  experimental searches for magnetic monopole particles have so far been unsuccessful,
 virtual  magnetic monopoles are found to emerge in momentum space of topological materials \cite{momentum,Niu,Niu2,Sinova}, parameter space of Berry phase theory \cite{Berry1},
 and in the synthetic magnetic field produced by a spinor Bose-Einstein condensate \cite{spinorBEC,spinorBEC2}.
Recently,  a classical analog to these elusive particles has emerged as topological excitations within pyrochlore spin ice systems \cite{spinice1,spinice2}.
The investigation of these  virtual magnetic monopoles not only  provides conclusive  evidence of the existence of Dirac magnetic monopoles
but also is very  crucial to understand the topological excitation properties in various physical systems.

In this Letter, we for the first time report that the  rogue waves (RWs) contain Dirac monopoles.
RWs are a kind of usual nonlinear excitations that have been observed experimentally in optical fibers \cite{Kibler}, water wave tanks \cite{water,ChabchoubH}, and plasma systems \cite{Bailung}. Because of its exotic characteristic of ``appearing from nowhere and disappearing without a trace''  \cite{AhkmediveAS-09,Dudley-14}, RW is believed to be the cause of many ocean disasters and therefore has attracted much attention \cite{RWReview,RWReview2}.
By investigating density zeros for RWs in an extended complex coordinate plane, we find that a $n$th-order RW
contains $n (n+1) $ pairs of monopoles with opposite charges, and that the collision of these monopoles and the reconnection of the corresponding vector field will lead to energy conversion from interaction energy to kinetic energy and
are responsible for the exotic property  of RWs. Important implications of our findings on other nonlinear waves such as  breathers or even some non-integrable excitations are  also discussed.

\emph{Dirac virtual magnetic monopole and topological vector potential}---We choose one of the simplest models, i.e., the scalar nonlinear Schr\"odinger equation (NLSE) $ i \frac{\partial \psi}{\partial t}= -\frac{1}{2} \frac{\partial^2 \psi}{\partial x^2} + g |\psi|^2 \psi$, to demonstrate our theory. NLSE can be  used to describe dynamics of nonlinear waves in atomic gases \cite{BEC}, plasma \cite{plasma}, water waves \cite{water}, and ferromagnetic materials \cite{ferro} and plays important roles in both integrability theory \cite{integrable} and optical communications \cite{optcomm}. It has a nonlinear wave solution $\psi(x,t)$, which can be a dark soliton (for $g>0$), a bright soliton, a RW, a breather (for $g<0$), or some superposition thereof. Neverthless, the topological properties associated with these one-dimensional (1D) nonlinear waves are seldom discussed \cite{Zhaoliu}.

 In fact, almost a century ago, P. A. M. Dirac has discussed the  phase change and its underlying topological property for a general wave function   $\psi(\textbf{r}, t)=|\psi| e^{i\phi}$ \cite{Dirac}. He introduced a vector potential field by the phase gradient, i.e, $\textbf{A}=(A_x,A_y,A_z)=\frac{\partial \phi}{\partial x} \textbf{e}_x+\frac{\partial \phi}{\partial y} \textbf{e}_y+\frac{\partial \phi}{\partial z} \textbf{e}_z$, and found that the noncommutative property of $\frac{\partial A_\mu}{ \partial \nu} \neq  \frac{\partial  A_\nu}{\partial \mu}$ ($\nu,\mu =x,y,z$ and $\nu\neq\mu$) would give rise to a non-integrable phase factor. This striking property is caused by the emergence  of the nodal line, where the wave function vanishes  and its phase does not have a meaning. More importantly, Dirac claimed that for calculating the change in phase round a closed curve, one has to take into account the influence of the end point of the nodal line, which acts as a virtual magnetic monopole of charge $\pm 1/2$ with a
 vector potential taking  the form of $\textbf{A}_{3D}=\frac{y \textbf{e}_x-x \textbf{e}_y}{2 r (r\pm z)}$ with $r= \sqrt{x^2+y^2+z^2}$ \cite{Dirac}.

However, when directly applying  the above Dirac's theory  to  the nonlinear waves such as RWs, we find that the non-integrable phase factor can not appear because the spatial coordinate is 1D so that  the noncommutative property has no meaning. It is noted that  some new phenomena would remain hidden if one were to restrict one's attention to real physical parameters \cite{Ananikian}. Two famous examples are the Lee-Yang zeros \cite{Lee-Yang} and Fisher zeros \cite{Fisher} reported for imaginary magnetic fields and imaginary temperatures, respectively. Very recently, the complex extension is also found to be helpful to study the fluid of Korteweg-de Vries equation \cite{Crabb}.
Here,  we attempt to  extend the real coordinate variable $x$ to a complex variable $Z =x+i y$, and  introduce a complex vector potential  also by  the phase gradient, i.e.,
\begin{eqnarray}
\mathscr{A}_{c} &=&\frac{\partial \phi(Z)} {\partial x} \textbf{e}_x+  \frac{\partial \phi(Z)} {\partial y} \textbf{e}_y\nonumber\\   &=&  \frac{\partial \phi(Z)} {\partial Z} (\textbf{e}_x+ i \textbf{e}_y). \label{eq1}
\end{eqnarray}
The vector potential is now defined on 2D plane of the variables of $x$ and $y$, and we focus on the real part of the above vector potential, i.e., $\textbf{A}=\rm{Re}[\mathscr{A}_{c}]$.
Following Dirac's spirit, we then pay attention to   the singularity property of the  vector potential field of  $\textbf{A}=(A_x,A_y)$ and its noncommutative property of $\frac{\partial A_x}{ \partial y} \neq  \frac{\partial  A_y}{\partial x}$. In the 2D situation, the nodal line will reduce to some scattered points corresponding to density zeros where the wave function vanishes  and its phase does not have a meaning. More interestingly, we  claim that for calculating the change in phase along $x$ axis, the density zeros (i.e., denoted by $\textbf{r}_{N}=x_N \textbf{e}_x +  y_N \textbf{e}_y$ ) of these nonlinear waves on an extended complex plane can constitute virtual magnetic monopole fields with a quantized flux of elementary $\Omega=\pi$, i.e., the  charge of $\mu=1/2$.
The corresponding vector potential takes  following explicit form:
\begin{eqnarray}
\textbf{A}=\rm{Re}[\mathscr{A}_c]= \sum_N \frac{ \pm \mu [(x-x_N) \textbf{e}_y-(y-y_N) \textbf{e}_x]} { (x-x_N)^2+(y-y_N)^2}, \label{eq2}
\end{eqnarray}
where $\pm$ is the sign of monopole. Near each singular point,  the above  2D expression (2) of  vector potential field 
is exactly the Dirac's monopole field of  $\textbf{A}_{3D}$ with setting $z=0$ therein. The corresponding magnetic field will be zero everywhere except at those singular points, that is, $\textbf{B}=\nabla \times \textbf{A}= \textbf{e}_z \sum\limits_N \pm \Omega \delta(x-x_N,y-y_N)$.

 The phase variations of such a nonlinear wave between $x_1$ and $x_2$ can then be expressed by the line integral of the vector potential, namely,
\begin{eqnarray}
\Delta\phi&=&\phi(x_2)-\phi(x_1)= \int_{x_1}^{x_2}  \textbf{A}(x,y=0) \cdot \textbf{e}_x d x \nonumber\\
    &=& \int_{(x_1,0)}^{(x_2,0)} \textbf{A}(x,y) \cdot d\vec{l} -\sum n \pi,  \label{eq3}
\end{eqnarray}
where $\vec{l}$ is an arbitrary curve that connects $(x_1,0)$ and $(x_2,0)$, and $\sum n \pi$ denotes the total magnetic flux enclosed by the curve $\vec{l}$ and the line connecting $x_1$ and $x_2$ in $(x,y)$ plane.

It is interesting to compare the above results with the Aharonov-Bohm effect \cite{AB}, which predicts a topological phase when an electron moves on a close path around a solenoid.
A 1D nonlinear wave moving on the real axis cannot see the magnetic fields scattered on the complex plane; however, it will acquire a phase  due to the presence of the vector potential.
The evolution of such a nonlinear wave can be understood from the transformed equation based on the topological vector potential, $i \partial_t \psi_1= (\frac{(\hat{p}_x+ \textbf{A}(x,y=0) \cdot \textbf{e}_x )^2}{2} +(g|\psi_1|^2 + \frac{\partial \int \textbf{A}(x,y=0) \cdot \textbf{e}_x d x }{\partial t} ) \psi_1$
with a transformation $\psi_1=\psi \exp{[-i \int \textbf{A}(x,y=0) \cdot \textbf{e}_x d x ]}$. The effective magnetic field and electric field can be derived as $\textbf{B}=\nabla \times \textbf{A}$ and $\textbf{E}=- \frac{\partial \textbf{A}}{\partial t} - \nabla  \frac{\partial \int \textbf{A}(x,y=0) \cdot \textbf{e}_x d x }{\partial t} $, respectively \cite{Jackiw}. In this sense, the phase variations of these nonlinear waves can be viewed as a 1D counterpart to the Aharonov-Bohm phase. Usually, the topological vector potential exhibits time dependence, which provides an alternative way to understand the dynamics of such nonlinear waves.

\begin{figure}[t]
\begin{center}
\includegraphics[width=8.5cm]{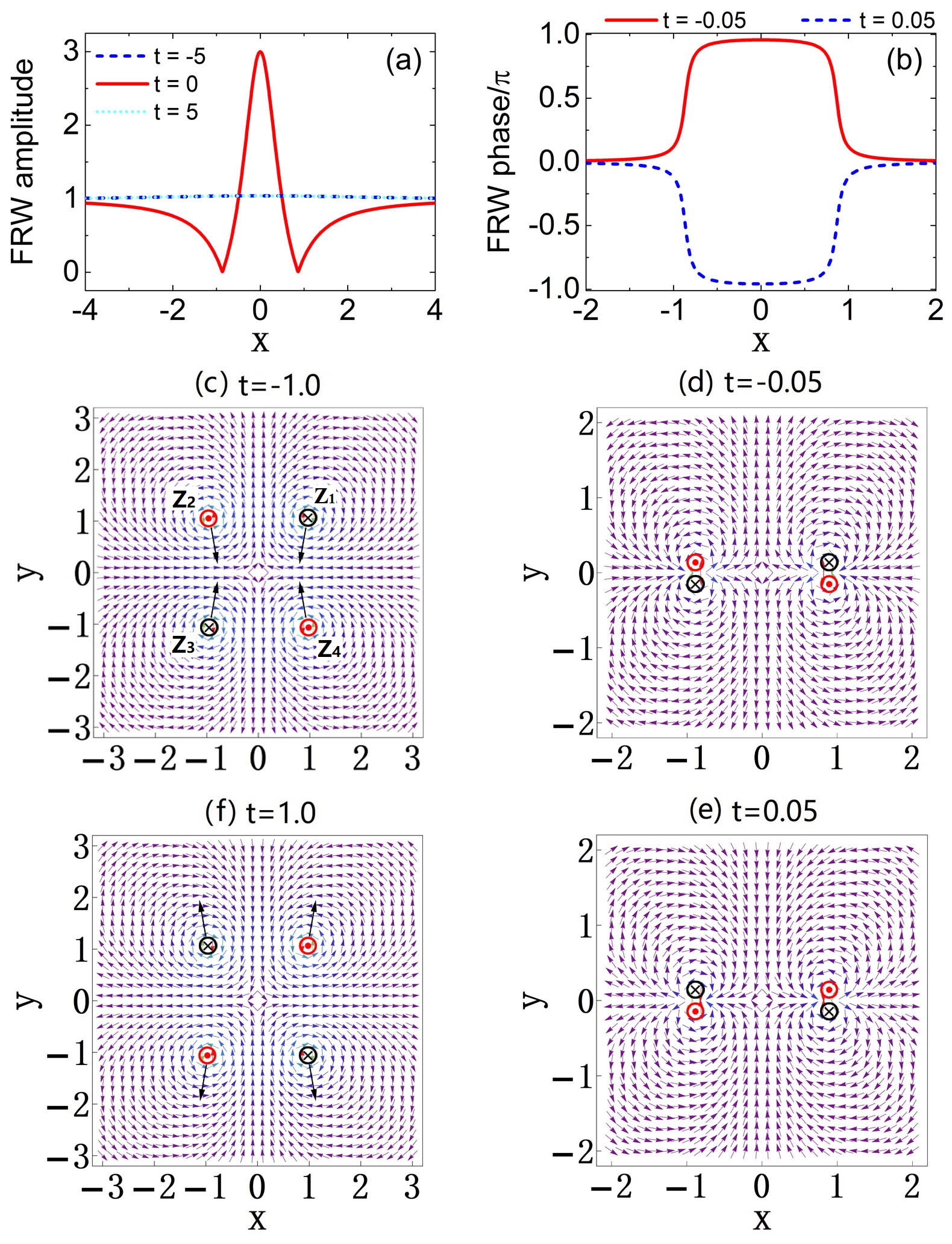}
\end{center}
\caption{ (a) Amplitude distribution, (b) phase distribution, and (c-f) evolution of the topological vector potential for a Peregrine breather (the first-order RW). The Dirac monopoles with positive and negative charges (i.e., $\pm \frac{1}{2}$) are indicated by $\bigodot$ and $\bigotimes$,  respectively. }\label{Fig1}
\end{figure}

\emph{Topological vector potentials for  rogue waves}---The above NLSE with $g=-1$ has a Peregrine breather which is also called as the first-order RW (FRW) solution on a uniform background \cite{Peregrine,AhkmediveAS-09,Ling}, $\psi = [1-\frac{4 (1+2 i t)}{4 t^2+4 x^2+1}] \ e^{i t}$. The temporal evolution of the RW amplitude depicted in Fig.~\ref{Fig1} (a) shows that the wave density remains almost constant until $t=-5$, after which sudden
growth occurs. At $t=0$, the amplitude amplification ratio  (defined as the peak
amplitude divided by the background amplitude) reaches its maximum value of $3$. Subsequently, the RW quickly decays, and the wave density recovers to be nearly constant. The amplitude peak is located at $x=0$, and on either side of the peak, there are two valleys at $x=\pm a_0$ ($a_0=\frac{\sqrt{3}}{2}$).
Interestingly, phase jumps accompany the rise in amplitude.
In Fig.~\ref{Fig1} (b), we see that there is a $\pi$ phase jump corresponding to each amplitude valley \cite{Akhmediev,Xu}, and the direction of the phase jump at $x=-a_0$
suddenly inverts to $-\pi$ slightly after the moment when the maximum amplitude peak emerges.

Linear modulational instability (MI) analysis describes well the growth of perturbation \cite{Dudley-14,RWReview,RWReview2,BaronioCDLOW-14,ZhaoL-16}. Nonlinear MI theory was further developed to address nonlinear behavior of MI beyond its growth stage \cite{NMI,NMI2}. A truncated three-wave model on frequency was suggested to describe well MI dynamics \cite{Trillo,Mussot,Gomel}.  Those could be used to understand the RW dynamics qualitatively.  Especially, analytical studies have indicated that the amplitude amplification ratios for RWs of different orders are subject to certain limits \cite{AhkmediveAS-09,Kedziora,He}, but no physical mechanism for these ceiling values has been discovered. The above $\pi$ phase jumps and abrupt inversion are also not fully understood \cite{Akhmediev,Xu}. Here, we attempt to elucidate these issues with the help of our developed topological vector potential theory.

\begin{figure}[t]
\begin{center}
\includegraphics[height=66mm,width=85mm]{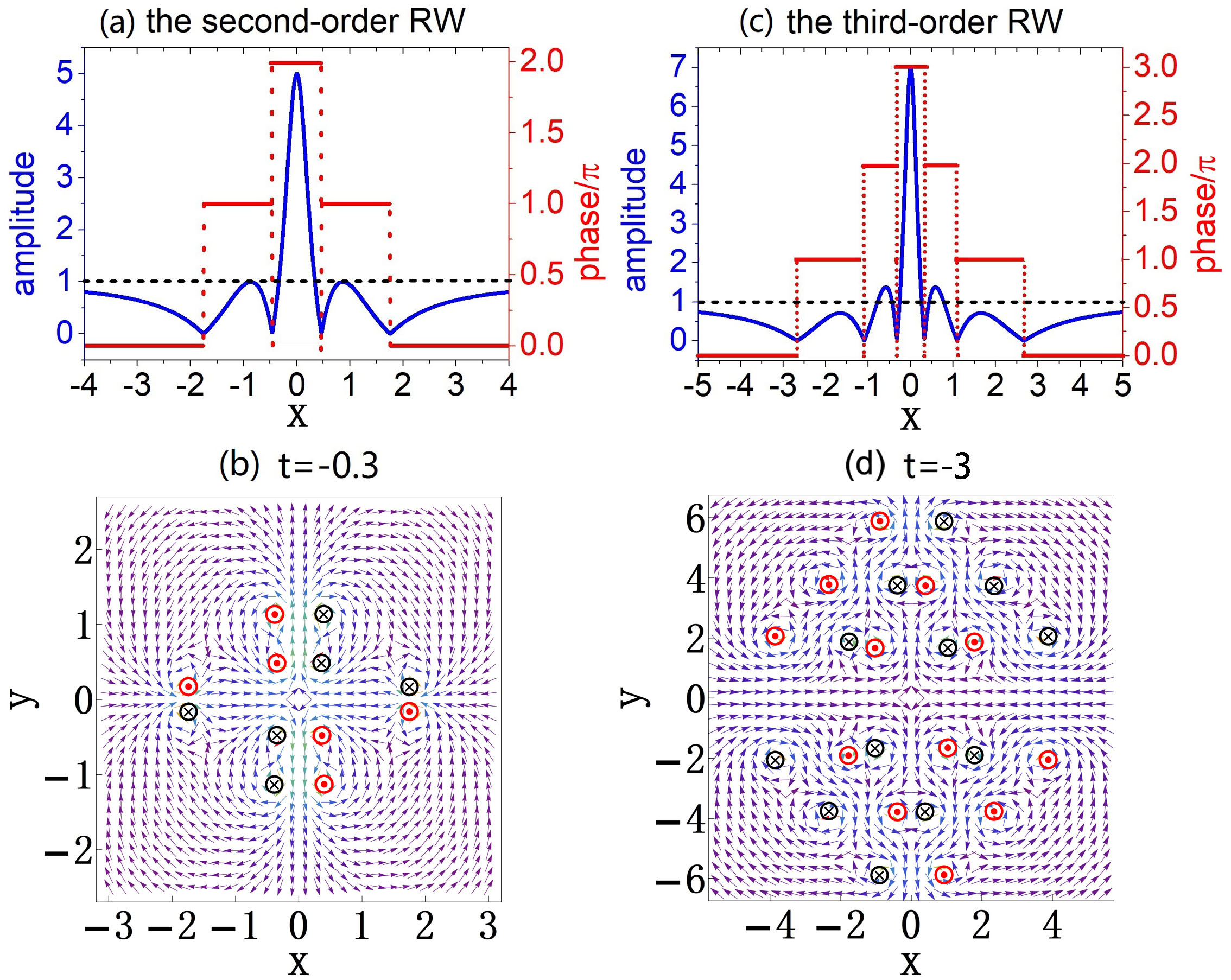}
\end{center}
\caption{(a) Amplitude and phase distributions for a second-order rogue wave (RW) at the moment when the wave amplitude reaches its highest peak ($t=0$). (b) Monopole vector potential at the moment slightly before the monopole collisions ($t=-0.3$). (c) Amplitude and phase distributions for a third-order RW at the moment when the wave amplitude reaches its highest peak ($t=0$). (d) Monopole vector potential at the moment before the Dirac monopole collisions ($t=-3$).
}\label{Fig2}
\end{figure}

It has four density zeros, i.e., $Z_{1,3}=\pm (a+i b)$ and $Z_{2,4}=\pm (-a+ib)$, where $a=\frac{2 \sqrt{2} |t|}{\sqrt{4 t^2+\sqrt{16 t^4+40 t^2+9}-3}}$ (the absolute operation in $|t|$ is introduced for marking monopole conveniently) and $b=\frac{\sqrt{4 t^2+\sqrt{16 t^4+40 t^2+9}-3}}{2 \sqrt{2}}$. According to Eq. (\ref{eq2}), the vector potential $\textbf{A}$ underlying the FRW takes an explicit form of  $\textbf{A}[x,y]= \sum\limits_{N=1,...,4} \frac{ \pm(-1)^N \mu   [(x-x_N) \textbf{e}_y-(y-y_N) \textbf{e}_x]} { (x-x_N)^2+(y-y_N)^2}$.
We can see that the vector potential $\textbf{A}$ is composed of two pairs of Dirac monopoles, and
in each pair, the two monopoles have charges of $1/2$ with opposite signs. The temporal evolution of the potential
is shown in Fig.~\ref{Fig1} (c-f).
The monopoles with opposite charges in each pair approach each other (see Fig.~\ref{Fig1} (c-d)), collide elastically in the vertical direction at time $t=0$ with speeds of $db/dt=\frac{2}{\sqrt{3}}$ and $da/dt=0$, and then bounce back after exchanging their charges (see Fig.~\ref{Fig1} (e-f) and a video in \cite{SPM}).
As $t\rightarrow \pm 0$, $a\rightarrow  a_0$ and $b\rightarrow 0$.
The vector potential on the real axis then takes the following form:
\begin{align}
\lim_{t\rightarrow \pm 0}\textbf{A}&=\lim_{b \rightarrow 0} \{\frac{-\mu [(x\pm a_0) \textbf{e}_y+ b \textbf{e}_x]} {(x\pm a_0)^2+ b^2]} + \frac{\mu [(x\pm  a_0) \textbf{e}_y- b \textbf{e}_x]} { (x\pm a_0)^2+b^2}  \nonumber\\
&+ \frac{\mu [(x\mp a_0) \textbf{e}_y+ b \textbf{e}_x]} {(x\mp a_0)^2+ b^2} + \frac{-\mu [(x\mp a_0) \textbf{e}_y- b \textbf{e}_x]} {(x\mp a_0)^2+b^2} \} \nonumber\\
&=[-\pi \delta(x\pm a_0)+\pi \delta(x\mp a_0)] \  \textbf{e}_x.
\end{align}
The line integral of the above vector potential can explain the $\pi$ phase jumps shown in Fig.~\ref{Fig1} (b).
The phase gradient determines the density flow, and the change in the phase distribution can provide an understanding of the growth and decay of RWs \cite{Akhmediev}.
The collision of the monopoles leads to a sudden rise in the wave amplitude. The exchange of the monopole charges after collision
can well explain the striking phase reversal that induces the RW's rapid decay. This provides another possible way to understand RW's dynamical properties, as a good supplement for the previously known mechanisms \cite{Dudley-14,Trillo,NMI}.

\begin{figure}[t]
\begin{center}
\includegraphics[height=38mm,width=85mm]{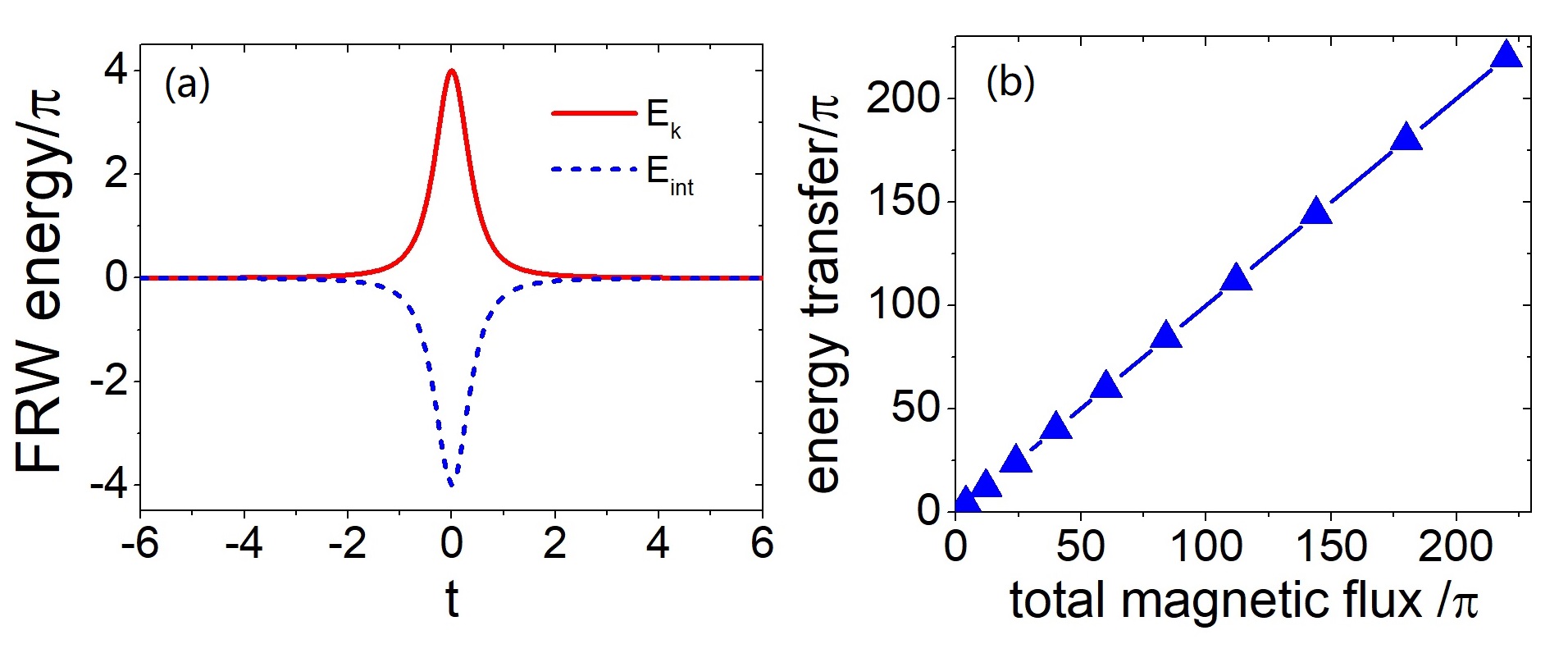}
\end{center}
\caption{(a) Temporal evolution of the kinetic energy and interaction energy of an FRW. (b)
Energy transfer vs. total magnetic flux. Here, the energy transfer refers to the time integral of the kinetic energy, and
the total magnetic flux represents the sum of the absolute magnetic fluxes of all monopoles for a high-order RW. The triangle is given by the calculations on RWs whose orders are up to $10$, and the solid line denotes the equal relation between energy transfer and the total magnetic flux. }\label{Fig3}
\end{figure}

Higher-order RWs admit higher amplitude peaks, more density valleys (or humps) and multiple phase steps,
as shown in Fig.~\ref{Fig2} \cite{AhkmediveAS-09,Akhmediev,Kedziora,Ling}. For a second-order RW, the maximum amplitude amplification ratio is $5$. There are five phase steps distributed symmetrically with respect to the $x=0$ axis, each of which is associated with a phase jump of $\pm \pi$ (see Fig.~\ref{Fig2} (a)). The vector field is composed of six pairs of Dirac monopoles, as shown in Fig.~\ref{Fig2} (b).
They can be divided into two classes: the four pairs that are closer to the x-axis collide with each other on the x-axis, leading to four $\pi$ phase jumps, whereas
the other two pairs (upper and lower pairs) collide on the imaginary axis, mainly contributing to a sudden rise in the wave amplitude. For a third-order RW, the maximum amplitude amplification ratio is $7$, and there are seven phase steps distributed symmetrically with respect to the $x=0$ axis (see Fig.~\ref{Fig2} (c)).
The vector field is composed of twelve pairs of monopoles,
as shown in Fig.~\ref{Fig2} (d). The paired monopoles collide and merge when the RW reaches its highest peak. Among them, six pairs collide and merge on the real axis, and the other six pairs collide on the imaginary axis or in other locations on the complex plane. The monopole pairs that do not collide on the real axis do not contribute to multiple phase jumps,
but they do influence the peak amplitude and the energy transfer involving RWs.

\emph{Energy transfer and topological vector potential reconnection}---The energy of a RW can be written as $\int_{-\infty}^{+\infty} [\frac{1}{2} |\partial_x\psi|^2- \frac{1}{2} (|\psi|^2-1)^2] dx$, where the first term is the kinetic energy $E_k$ and the second term is the interaction energy $E_{int}$.
For an FRW, the time-dependent kinetic energy is
$E_k =\frac{4 \pi }{(4 t^2+1)^{3/2}}.$
Because the total energy is conserved, the amplitude amplification of the RW corresponds to the energy transfer process from interaction energy to kinetic energy, as indicated in Fig.~\ref{Fig3} (a).
Quantitatively, the total energy transfer can be evaluated as the time integral of the kinetic energy ($\int_{-\infty}^{+\infty} E_k dt$). We have numerically calculated the integrals for RWs whose orders are up to $10$ and have found that they are equal
to the sum of the absolute magnetic flux of the monopoles (see Fig.~\ref{Fig3} (b)).
From the vector field perspective, we know that monopole collisions induce the conversion of interaction energy into kinetic energy. This process is analogous to the magnetic field reconnection process identified in astrophysics, in which magnetic field energy is transformed into the kinetic energy of a plasma \cite{magnerecon}.

We have also investigated the relation between the peak amplitude of a RW and the number of monopoles and found that they are closely related.
According to our discussion above, a FRW admits $4$ Dirac monopoles.
Because the $n$th-order RW solution is a nonlinear superposition of $\frac{n (n+1)} {2}$ FRWs \cite{Kedziora},
it will contain $N= 2 n (n+1) $ monopoles on the complex plane.
Among them, there are $4n$ monopoles colliding on the real axis that are responsible for the multiple phase steps ($n$th-order RW with the highest peak), whereas the other $2n (n-1)$ monopoles will collide in other locations on the complex plane. On the other hand, the square of the maximum amplitude amplification ratio $P$ for a $n$th-order RW can be calculated to be $(2 n+1)^2$ \cite{AhkmediveAS-09,He}. We thus obtain the following explicit relation: $P=2 N+1$. Based on this observation, we can predict that the amplitude amplification ratio of a high-order RW will be subject to a certain limit because a RW contains only a limited number of Dirac monopoles due to the finite number of valleys in its geometric configuration \cite{Ling,Kedziora}.

\emph{Conclusion and discussion}---With extending to complex coordinate plane and following Dirac's theoretical formulation of virtual magnetic monopole, we reveal a monopole topological potential hidden in the phase variations of 1D nonlinear waves.  The monopole collision process can be used to understand the striking phase jump of RWs. Moreover, we find that the ceiling values of the  amplitude amplification ratios for RWs of different orders are determined by the total magnetic flux. The unintelligible $\pi$ phase jump is found to be the manifestation of the quantized magnetic flux  of a magnetic monopole. Our finding provides a distinct explanation for the exotic property of ``appearing from nowhere and disappearing without a trace '' of RWs. This could be a good supplement for the previously known mechanisms \cite{RWReview,RWReview2}.

Our theory is also applicable to other nonlinear waves \cite{Kivshar,AhB,K-M,BS} (see details in \cite{SPM}) and nonlinear models \cite{CQds,Chen,DBS,Zhao,Liu,zhaoliu,DBB,Qds}. It can be further extended to study the semiclassical dynamics of BdG excitations in Bose-Einstein condensation under the action of intrinsic topological potentials \cite{Zhang}, and the interaction between nonlinear waves even in non-integrable systems \cite{Armaroli}. In the latter case, the density zeros in the complex plane that constitute the magnetic monopole fields could be traced with the help of numerical algorithm.

\emph{Acknowledgments}
We thank Prof. C. H. Lee for noticing the connection to Lee-Yang zeros. This work was supported by the National Natural Science Foundation of China (Contract No. 12022513, 11775176, 12047502 and 11775030), NSAF No. U1930403, and the Major Basic Research Program of Natural Science of Shaanxi Province (Grant No. 2018KJXX-094, 2017KCT-12).


\begin{thebibliography}{99}
\bibitem{MMr} A. S. Goldhaber,  W. P. Trower, (eds) Magnetic Monopoles (American Association
of Physics Teachers, 1990).
\bibitem{MMr2} K. A. Milton, Theoretical and experimental status of magnetic monopoles, \newblock\href{https://doi.org/10.1088/0034-4885/69/6/R02} {Rep. Prog. Phys. \textbf{69}, 1637-1711 (2006)}.
\bibitem{Dirac} P. A. M. Dirac, Quantised singularities in the electromagnetic field, \newblock\href{https://doi.org/10.1098/rspa.1931.0130} {Proc. R. Soc. Lond. A \textbf{133}, 60-72 (1931)}.
\bibitem{Rebbi} C. Rebbi,  G. Soliani  (eds), Solitons and particles (Singapore: World Scientific Publishing, 1985).
\bibitem{Hooft} G.t Hooft,  Magnetic monopoles in unified gauge theories, \newblock\href{https://doi.org/10.1016/0550-3213(74)90486-6} {Nucl. Phys. B \textbf{79}, 276-284 (1974)}.
\bibitem{sff}  H. Georgi and S. Glashow, Unity of All Elementary Particle Forces, \newblock\href{https://doi.org/10.1103/PhysRevLett.32.438} {Phys. Rev. Lett. \textbf{32},
438 (1974)}.
\bibitem{momentum} Z. Fang,  N. Nagaosa, Kei S. Takahashi, A. Asamitsu, R. Mathieu,  T. Ogasawara,  H. Yamada,  M. Kawasaki,  Y. Tokura, K. Terakura, The anomalous Hall effect and magnetic monopoles in momentum
space,  \newblock\href{https://doi.org/10.1126/science.1089408} {Science \textbf{302}, 92-95 (2003)}.
\bibitem{Niu} D. Xiao, M.-C. Chang, and Q. Niu, Berry phase effects on electronic properties,
\newblock\href{https://doi.org/10.1103/RevModPhys.82.1959} {Rev. Mod. Phys. \textbf{82}, 1959 (2010)}.
\bibitem{Niu2} Z. Qiao, H. Jiang, X. Li, Y. Yao, and Q. Niu, Microscopic theory of quantum anomalous Hall effect in graphene, \newblock\href{https://doi.org/10.1103/PhysRevB.85.115439}{Phys. Rev. B \textbf{85}, 115439 (2012).}
\bibitem{Sinova} J. Sinova, D. Culcer, Q. Niu, N. A. Sinitsyn, T. Jungwirth, and A. H. MacDonald, Universal Intrinsic Spin Hall Effect, \newblock\href{https://doi.org/10.1103/PhysRevLett.92.126603}{Phys. Rev. Lett. \textbf{92}, 126603 (2004).}
\bibitem{Berry1} M.V. Berry, Quantal phase factors accompanying adiabatic changes,  \newblock\href{https://doi.org/10.1098/rspa.1984.0023} {Proc. R. Soc. Lond. A \textbf{392}, 45-57 (1984)}.
\bibitem{spinorBEC} V. Pietila,  M. Mottonen, Creation of Dirac monopoles in spinor Bose-Einstein
condensates, \newblock\href{https://doi.org/10.1103/PhysRevLett.103.030401} {Phys. Rev. Lett. \textbf{103}, 030401 (2009)}.
\bibitem{spinorBEC2} M. W. Ray, E. Ruokokoski, S. Kandel, M. Mottonen, D. S. Hall, Observation of Dirac monopoles in a synthetic
magnetic field, \newblock\href{https://doi.org/10.1038/nature12954} {Nature \textbf{505}, 657 (2014)}.

\bibitem{spinice1} C. Castelnovo,  R. Moessner,  S. L. Sondhi, Magnetic monopoles in spin ice, \newblock\href{https://doi.org/10.1038/nature06433} {Nature \textbf{451}, 42-45 (2008)}.
\bibitem{spinice2} D. J. P.  Morris, D. A. Tennant, S. A. Grigera, B. Klemke, C. Castelnovo, R. Moessner, C. Czternasty, M. Meissner, K. C. Rule, J.-U. Hoffmann,  Dirac strings andmagneticmonopoles in the spin ice Dy2Ti2O7,
\newblock\href{https://doi.org/10.1126/science.1178868} {Science \textbf{326}, 411-414 (2009)}.

 \bibitem{Kibler} B. Kibler, J. Fatome, C. Finot, G. Millot, F. Dias, G. Genty, N. Akhmediev, and J. M. Dudley, The Peregrine soliton in nonlinear fibre optics, \newblock \href{https://doi.org/10.1038/nphys1740} {Nature Phys. \textbf{6}, 790-795 (2010).}
\bibitem{water} A. Chabchoub, N. P. Hoffmann, and N. Akhmediev, Rogue wave observation in a water wave tank, \newblock\href{https://doi.org/10.1103/PhysRevLett.106.204502}{Phys. Rev. Lett. \textbf{106}, 204502 (2011).}
\bibitem{ChabchoubH} A. Chabchoub, N. Hoffmann, M. Onorato, A. Slunyaev, A. Sergeeva, E. Pelinovsky, and N. Akhmediev, Observation of a hierarchy of up to fifth-order rogue waves in a water tank, \newblock \href{https://doi.org/10.1103/PhysRevE.86.056601}{Phys. Rev. E \textbf{86}, 056601 (2012).}
\bibitem{Bailung} H. Bailung, S. K. Sharma, and Y. Nakamura, Observation of Peregrine Solitons in a Multicomponent Plasma with Negative Ions, \newblock \href{https://doi.org/10.1103/PhysRevLett.107.255005}{Phys. Rev. Lett. \textbf{107}, 255005 (2011).}


\bibitem{AhkmediveAS-09} N. Akhmediev, A. Ankiewicz, and J. M. Soto-Crespo, Rogue waves and rational solutions of the nonlinear Schr\"odinger equation, \newblock\href{https://doi.org/10.1103/PhysRevE.80.026601} {Phys. Rev. E  \textbf{80},  026601 (2009).}
\bibitem{Dudley-14} J. M. Dudley, F. Dias, M. Erkintalo, and G. Genty, Instabilities, breathers and rogue waves in optics, \newblock\href{https://doi.org/10.1038/nphoton.2014.220}
{Nature Photon  \textbf{8}, 755-764 (2014).}
\bibitem{RWReview} C. Kharif, E. Pelinovsky, Physical mechanisms of the rogue wave phenomenon, \newblock\href{https://doi.org/10.1016/j.euromechflu.2003.09.002} {Euro. J.  Mech. B  \textbf{22}, 603-634 (2003).}
\bibitem{RWReview2} M. Onorato, S. Residori, U.Bortolozzo, A.Montina, F.T.Arecchi,  \newblock\href{https://doi.org/10.1016/j.physrep.2013.03.001} {Phys. Rep. \textbf{528}, 47-89 (2013).}


\bibitem{BEC} S. Burger, K. Bongs, S. Dettmer, W. Ertmer, K. Sengstock, A. Sanpera, G. V. Shlyapnikov, and M. Lewenstein,  Dark solitons in Bose-Einstein condensates, \newblock\href{https://doi.org/10.1103/PhysRevLett.83.5198}{Phys. Rev. Lett. \textbf{ 83}, 5198 (1999).}
\bibitem{plasma}  R. Heidemann, S. Zhdanov, R. S\"{u}tterlin, H. M. Thomas, and G. E. Morfill, Dissipative dark soliton in a complex plasma, \newblock\href{https://doi.org/10.1103/PhysRevLett.102.135002}{Phys. Rev. Lett. \textbf{102}, 135002 (2009).}

\bibitem{ferro}  W. Tong, M. Wu, L. D. Carr, and B. A. Kalinikos, Formation of random dark envelope solitons from incoherent waves, \newblock\href{https://doi.org/10.1103/PhysRevLett.104.037207}{Phys. Rev. Lett. \textbf{104}, 037207 (2010).}
 \bibitem{integrable} V. B. Matveev and M. A. Salle, Darboux Transformation
and Solitons (Springer-Verlag, Berlin, 1991); J. K. Yang,  Nonlinear Waves
in Integrable and Nonintegrable Systems (SIAM, Philadelphia, 2010).
 \bibitem{optcomm} G. P. Agrawal, Nonlinear Fiber Optics, 4th ed. (Acdemic, 2007).
 \bibitem{Zhaoliu} L.-C. Zhao, Y.-H. Qin, C. Lee, and J. Liu, Classification of Dark Solitons via Topological Vector Potentials,  \newblock
 \href{https://doi.org/10.1103/PhysRevE.103.L040204}%
 {Phys. Rev. E \textbf{103}, L040204  (2021).}

\bibitem{Ananikian} N. Ananikian,  R. Kenna, Imaginary Magnetic Fields in the Real World, \newblock\href{https://doi.org/10.1103/Physics.8.2} {Physics \textbf{8}, 2 (2015)}.
\bibitem{Lee-Yang}  C. N. Yang and T. D. Lee,  Statistical Theory of Equations of State and Phase Transitions. I. Theory of Condensation,
 \newblock\href{https://doi.org/10.1103/PhysRev.87.404}{Phys. Rev. \textbf{87}, 404 (1952);} T. D. Lee and C. N. Yang,  Statistical Theory of Equations of State and Phase Transitions. II. Lattice Gas and Ising Model,  \newblock\href{https://doi.org/10.1103/PhysRev.87.410}{Phys. Rev. \textbf{87}, 410 (1952).}
\bibitem{Fisher} M. E. Fisher in Statistical Physics and Solid State Physics, Lectures in Theoretical Physics, edited by W. E. Brittin (University of Colorado Press, Boulder, 1965).
\bibitem{Crabb} M. Crabb,  N. Akhmediev, Complex Korteweg-de Vries equation: A deeper theory of shallow water waves, \newblock
    \href{https://doi.org/10.1103/PhysRevE.103.022216} {Phys. Rev. E \textbf{103},  022216 (2021).}


\bibitem{AB} Y. Aharonov and D. Bohm, Significance of Electromagnetic Potentials in the Quantum Theory, \newblock
    \href{https://doi.org/10.1103/PhysRev.115.485} {Phys. Rev. \textbf{115}, 485 (1959).}

\bibitem{Jackiw} R. Jackiw and S.-Y. Pi, Soliton solutions to the Gauged Nonlinear Schr\"odinger equation on the plane, \newblock
    \href{https://doi.org/10.1103/PhysRevLett.64.2969} {Phys. Rev. Lett. \textbf{64}, 2969 (1990).}

\bibitem{Peregrine} D. H. Peregrine, Water waves, nonlinear Schr\"odinger equations and their solutions,
\newblock\href{ https://doi.org/10.1017/S0334270000003891} {J. Aust. Math. Soc. Ser. B \textbf{25}, 16-43 (1983).}

\bibitem{Ling} B. L. Guo, L. M. Ling, and Q. P. Liu, Nonlinear Schr\"{o}dinger equation:
generalized Darboux transformation and rogue wave solutions, \newblock\href{https://doi.org/10.1103/PhysRevE.85.026607} {Phys. Rev. E \textbf{85}, 026607 (2012).}
 \bibitem{Akhmediev} D. J. Kedziora, A. Ankiewicz, and N. Akhmediev, The phase patterns of higher-order rogue waves, \newblock \href{https://doi.org/10.1088/2040-8978/15/6/064011} {J. Opt. \textbf{15}, 064011 (2013).}
  \bibitem{Xu}  G. Xu, K. Hammani, A. Chabchoub, J. M. Dudley, B. Kibler, and C. Finot, Phase evolution of Peregrine-like breathers in optics and hydrodynamics, \newblock \href{https://doi.org/10.1103/PhysRevE.99.012207}{Phys. Rev. E \textbf{99}, 012207 (2019).}
\bibitem{BaronioCDLOW-14} F. Baronio, M. Conforti, A. Degasperis, S. Lombardo, M. Onorato, and S. Wabnitz, Vector rogue waves and baseband modulation instability in the defocusing regime, \newblock\href{https://doi.org/10.1103/PhysRevLett.113.034101} {Phys. Rev. Lett. \textbf{113}, 034101 (2014).}
\bibitem{ZhaoL-16} L.-C. Zhao and L. Ling, Quantitative relations between modulational instability and several well-known nonlinear excitations, \newblock\href{https://doi.org/10.1364/JOSAB.33.000850}{J. Opt. Soc. Am. B   \textbf{33}, 850-856 (2016).}
\bibitem{NMI} A. A. Gelash, and V. E. Zakharov, Superregular solitonic solutions: a novel scenario for the nonlinear stage of modulation instability, \newblock\href{https://doi.org/10.1088/0951-7715/27/4/R1} {Nonlinearity \textbf{27}, R1-R39 (2014).}
\bibitem{NMI2} S. Trillo, C. Naveau, P. Szriftgiser, M. Conforti, A. Kudlinski, F. Copie, and A. Mussot, Nonlinear Modulational Instability: Recurrences, Broken Symmetry, and Breathers, in Nonlinear Optics (NLO), OSA Technical Digest (Optical Society of America, 2019), \newblock\href{https://doi.org/10.1364/NLO.2019.NTu2A.6} {paper NTu2A.6.}
\bibitem{Trillo} S. Trillo and S. Wabnitz, Dynamics of the nonlinear
modulational instability in optical fibers, \newblock\href{https://doi.org/10.1364/ol.16.000986}{Opt. Lett.   \textbf{16}, 986-988 (1991).}
\bibitem{Mussot} A. Mussot,  C. Naveau, M. Conforti,  A. Kudlinski, F. Copie, P. Szriftgiser, S. Trillo, Fibre multi-wave mixing combs reveal the broken symmetry of Fermi-Pasta-Ulam recurrence, \newblock\href{https://doi.org/10.1038/s41566-018-0136-1} {Nature Photon. \textbf{12}, 303-s308 (2018)}.
\bibitem{Gomel} A. Gomel, A. Chabchoub, M. Brunetti, S. Trillo,
J. Kasparian, and A. Armaroli, Stabilization of Unsteady Nonlinear Waves by Phase-Space Manipulationf, \newblock\href{https://doi.org/10.1103/PhysRevLett.126.174501}{Phys. Rev. Lett. \textbf{126}, 174501 (2021)}.

\bibitem{Kedziora} D. J. Kedziora, A. Ankiewicz, and N. Akhmediev, Classifying the hierarchy of nonlinear-Schr\"{o}dinger-equation rogue-wave solutions,  \newblock\href{https://doi.org/10.1103/PhysRevE.88.013207} {Phys. Rev. E \textbf{88}, 013207 (2013).}
\bibitem{He} L. Wang, C. Yang, J. Wang, J.  He, The height of an nth-order fundamental rogue wave for the nonlinear Schr\"{o}dinger equation,  \newblock
    \href{https://doi.org/10.1016/j.physleta.2017.03.023} {Phys. Lett. A \textbf{381}, 1714-1718 (2017).}
\bibitem{SPM} The supplemental metarial for this paper.
\bibitem{magnerecon} E. Priest and  T. Forbes, Magnetic reconnection: MHD Theory and Applications (Cambridge University Press, New York, 2000).

\bibitem{Kivshar} Y. S. Kivshar and B. Luther-Davies, Dark optical solitons: physics and applications, \newblock\href{https://doi.org/10.1016/S0370-1573(97)00073-2}{Phys. Rep. \textbf{298}, 81 (1998).}

\bibitem{AhB} N. Akhmediev and V. I. Korneev, Modulational instability and periodic solutions of the
Nonlinear Schr\"{o}dinger equation, \newblock
    \href{https://link.springer.com/content/pdf/10.1007/BF01037866.pdf} {Theor. Math. Phys. \textbf{69}, 1089 (1986).}
\bibitem{K-M}  E. Kuznetsov, Solitons in a parametrically unstable plasma, Sov. Phys. Dokl. \textbf{22}, 507 (1977); Y. C. Ma, The Perturbed Plane-Wave Solutions of the Cubic Schr\"{o}dinger Equation, \newblock
    \href{https://doi.org/10.1002/sapm197960143} {Stud. Appl. Math. \textbf{60}, 43 (1979).}
\bibitem{BS} V. E. Zakharov and A. B. Shabat,   Exact theory of two-dimensional self-focusing and one-dimensional self-modulation of waves in nonlinear media, \newblock
    \href{http://zakharov75.itp.ac.ru/static/local/zve75/zakharov/1972/1972-05-e_034_01_0062.pdf} {Zh. Eksp. Teor. Fiz. \textbf{61}, 118-134 (1971).}

\bibitem{CQds} Y. S. Kivshar, V. V. Afansjev, and A. W. Snyder, Dark-like bright solitons, \newblock
    \href{https://doi.org/10.1016/0030-4018(96)00111-3} {Opt. Commun. \textbf{126}, 348 (1996).}

\bibitem{Chen} H. Triki, Y. Hamaizi, Q. Zhou, A. Biswas, M. Z. Ullah, S. P. Moshokoa, and M. Belic, Chirped dark and gray solitons for Chen-Lee-Liu equation in optical fibers and PCF, \newblock
    \href{https://doi.org/10.1016/j.ijleo.2017.11.038}  {Optik \textbf{155}, 329 (2018).}

\bibitem{DBS} Th. Busch and J.R. Anglin, Dark-Bright Solitons in Inhomogeneous Bose-Einstein Condensates, \newblock
    \href{https://doi.org/10.1103/PhysRevLett.87.010401} {Phys. Rev. Lett. \textbf{87}, 010401 (2001).}
\bibitem{Zhao} L.-C. Zhao, S.-C. Li, and L. Ling, Rational W-shaped solitons on a continuous-wave background in the Sasa-Satsuma equation, \newblock
    \href{https://doi.org/10.1103/PhysRevE.89.023210} {Phys. Rev. E \textbf{89}, 023210 (2014)}.
\bibitem{Liu} C. Liu, Z.-Y. Yang, L.-C. Zhao, and W.-L. Yang, State transition induced by higher-order effects and background frequency, \newblock
    \href{https://doi.org/10.1103/PhysRevE.91.022904} {Phys. Rev. E \textbf{91}, 022904 (2015)}.
\bibitem{zhaoliu}  L.-C. Zhao, W. Wang, Q. Tang, Z.-Y. Yang, W.-L. Yang, and J. Liu,
 \newblock Spin soliton with a negative-positive mass transition, \newblock
 \href{https://doi.org/10.1103/PhysRevA.101.043621}%
 {Phys. Rev. A \textbf{101}, 043621 (2020).}
 \bibitem{DBB} R. Radhakrishnan, N. Manikandan, and K. Aravinthan, Energy-exchange collisions of dark-bright-bright vector solitons, \newblock
    \href{https://doi.org/10.1103/PhysRevE.92.062913} {Phys. Rev. E \textbf{92}, 062913 (2015).}
\bibitem{Qds} M. A. Alejo and  A. J. Corcho, Orbital stability of the black soliton for the quintic Gross-Pitaevskii equation, \newblock
    \href{https://arxiv.org/abs/2003.09994} {arXiv:2003.09994 (2020).}

\bibitem{Zhang} C. Zhang, A.M. Dudarev, Q. Niu, Berry phase effects on dynamics of quasiparticles in a superfliud with a vortex, \newblock
    \href{https://doi.org/10.1103/PhysRevLett.97.040401} {Phys. Rev. Lett. \textbf{97}, 040401 (2006)}.
\bibitem{Armaroli} A. Armaroli,   C. Conti, F. Biancalana, Rogue solitons in optical fibers: a dynamical process in a complex energy landscape,  \newblock
    \href{http://dx.doi.org/10.1364/OPTICA.2.000497} {Optica \textbf{2},  497 (2015)}.
\end{thebibliography}
\end{document}